# Data description and retrieval using periods represented by uncertain time intervals

## TATSUKI SEKINO[1, a)]

**Abstract**: Time periods are frequently used to specify time in metadata and retrieval. However, it is not easy to describe and retrieve information about periods, because the temporal ranges represented by periods are often ambiguous. This is because these temporal ranges do not have fixed beginning and end points. To solve this problem, basic logics to describe and process uncertain time intervals were developed in this study. An uncertain time interval is represented as a set of time intervals that indicate states when the uncertain time interval is determined. Based on this concept, a logic to retrieve uncertain time intervals satisfying a given condition was established, and it was revealed that retrieval results belong to three states: reliable, impossible, and possible matches. Additionally, to describe data about uncertain periods, an ontology (the HuTime Ontology) was constructed based on the logic. This ontology is characterized by the fact that uncertain time intervals can be defined recursively. It is expected that more data about time periods will be created and released using the result of this study.

## 1. Introduction

Time is often expressed using periods. "Renaissance," "Industrial Revolution," and "Cold War," which are historical eras or events, are some typical examples. These periods are used not only in sentences, but also in metadata such as the periods when objects (for example, persons, documents, and events) exist. However, criteria to apply these periods in metadata are ambiguous. For example, a painting may be classified as a Renaissance work by one researcher but may be classified as a non-Renaissance work by another researcher, even though its year of creation is clear. This problem is due to the ambiguity of the temporal range indicated by the Renaissance. The Renaissance period naturally does not have fixed beginning and end points, because it did not start or end suddenly on one day. Therefore, the temporal range of the Renaissance is usually represented as a period that has ranges of beginning and end points, such as "from the 14th century to the 16th century."

Temporal information is effectively processed on computer systems, and used for visualization, as timelines and charts and for various kinds of analysis, in addition to retrievals. However, most computer systems assume that temporal data are given as determinate values of year-month-day, and do not accept uncertain periods that have the beginning and end point as ranges. Consequently, data with uncertain periods have not been used effectively, although historical eras and events represented by uncertain periods are widely used in metadata and documents.

In this study, we construct logics to process uncertain periods, and create resources to implement the logics. Concretely, we consider logics to express and compare uncertain time intervals, focusing on realizing retrievals of uncertain periods, and then we create an ontology and vocabularies to describe data about uncertain time intervals, based on the logics. Finally, we verify that the logics and the resources are actually implementable, using experimental implementation.

## 2. Related Studies

Recently, some international standards relating to time have been released, and interest in temporal information is increasing. In 2017, OWL-Time was released as a W3C recommendation [1]. OWL-Time is an ontology that can process temporal information in the framework of OWL, which is an ontology for semantic web technology [2]. In 2019, ISO 8601-2 was released [3]. ISO 8601-2 is an extension of ISO 8601, which has long been widely used as a standard format to represent date and time [4], and provides an extended format to represent calendar periods such as year, month, and decade (e.g., the 2010s). The release of ISO 8601-2 indicates the demand for processing uncertain periods, which cannot be represented by the conventional ISO 8601. These international standards will contribute to the description of temporal information and will enhance studies that use temporal information. However, neither these standards nor other existing standards contain the logics or procedures required for retrievals and analyses using uncertain periods.

Some databases have a function whereby users can specify temporal conditions for retrieval using periods such as historical eras or events. However, no standards for the temporal ranges indicated by the periods have been established. A major concern is that temporal ranges indicated by the periods may differ between databases, even if the names of the periods are the same. To solve this issue, some projects have constructed a thesaurus, which collects information about various types of periods, and associates them with an explanation including a description of the temporal range (e.g., DBPedia [5] and Art & Architecture Thesaurus [6]). Furthermore, there are services to provide information about periods, including uncertain periods, represented by ranges of the beginning and end points (e.g., PeriodO [7]). However, practical procedures to compare between uncertain periods have not been established even though it is required for retrievals. Consequently, these resources have not

---

been used effectively in databases, though they are available to describe temporal data.

Various logics relating to time intervals representing uncertain periods have been proposed [8]. Although these logics attempt to represent uncertain time intervals using fuzzy sets [9], rough sets [10], or both [11], most of them are limited to proposing the logics, and there are fewer examples of practical implementation. Additionally, most of the studies have focused only on representing uncertain periods, and logics to examine the temporal relations between them are insufficient.

This study aims to address the lack of logics and resources for realizing the retrieval of uncertain time intervals. Sekino (2019) [12] proposed the concept that uncertain time intervals are represented as sets of determinate time intervals, and constructed a logic to compare them, based on Allen's relations [13]. Although this concept is useful for the purpose of this study, it is insufficient for actual retrieval, because it does not sufficiently consider procedures to compare uncertain time intervals with a given retrieval condition or logics to process time instants. In this study, we construct new logics and procedures to represent and compare uncertain time intervals, based on the concept of Sekino (2019).

## 3. Logics to process uncertain time intervals

In this section, we construct logics to realize retrieval using uncertain time intervals. Sections 3-1 and 3-2 are basic logics to process uncertain time intervals; they reinforce and improve on the concept of Sekino (2019). Sections 3-3 and 3-4 are logics that are newly constructed for the purpose of this study.

### 3.1 Uncertain time intervals

A time interval between a pair of time instants is called a *determinate time interval*, denoted as $\hat{\omega}$. A determinate time interval is a continuous interval of real numbers, and its boundaries—the beginning and end points ($\hat{\omega}_b$ and $\hat{\omega}_e$)—are not uncertain (Equation 1).

$$\hat{\omega} := \{x \in \mathbb{R} | \hat{\omega}_b \leq x \leq \hat{\omega}_e\} \tag{1}$$

When one or both of the boundaries of a time interval is not fixed to a specific value, this boundary can be represented as a temporal range. This means that a time interval whose beginning and end points are represented by ranges contains two or more determinate time intervals. That is, any time interval with two or more different determinate states is uncertain. In this study, we call a set of determinate time intervals an *uncertain time interval*, denoted as $\omega$ (Equation 2).

$$\omega := \{\hat{\omega} | \omega_{Pb} \leq \hat{\omega}_b \leq \omega_{Rb} \wedge \omega_{Re} \leq \hat{\omega}_e \leq \omega_{Pe}\} \tag{2}$$

The ranges of the beginning and end points of the uncertain time interval are represented as between $\omega_{Pb}$ and $\omega_{Rb}$ and between $\omega_{Re}$ and $\omega_{Pe}$, respectively (Figure 1). We assume that these ranges are continuous intervals, to simplify the logic.

We can consider the summation of the temporal ranges of all determinate time intervals $\hat{\omega}$ contained in an uncertain time interval $\omega$. We define this summation of temporal ranges as a *possible time interval*, denoted as $\omega_P$ (Equation 3).

$$\omega_P := \bigcup_{\hat{\omega} \in \omega} \hat{\omega} \ \Rightarrow \ \omega_P = \{x | \omega_{Pb} \leq x \leq \omega_{Pe}\} \tag{3}$$

The beginning and end points of a possible time interval are $\omega_{Pb}$ and $\omega_{Pe}$, respectively, and $\omega_{Pb}$ is always less than or equal to $\omega_{Pe}$. The possible time interval indicates the temporal range that may be included in the uncertain time interval $\omega$. Therefore, outside of the possible time interval are temporal ranges that can never be included in the uncertain time interval $\omega$.

In contrast, we can consider a common temporal range for all determinate time intervals $\hat{\omega}$ contained in an uncertain time interval $\omega$. We define this common temporal range as a *reliable time interval*, denoted as $\omega_R$ (Equation 4).

$$\omega_R := \bigcap_{\hat{\omega} \in \omega} \hat{\omega}$$
$$\Rightarrow \begin{cases} \omega_{Rb} \leq \omega_{Re} \Rightarrow \omega_R = \{x | \omega_{Rb} \leq x \leq \omega_{Re}\} \\ \omega_{Rb} > \omega_{Re} \Rightarrow \omega_R = \emptyset \end{cases} \tag{4}$$

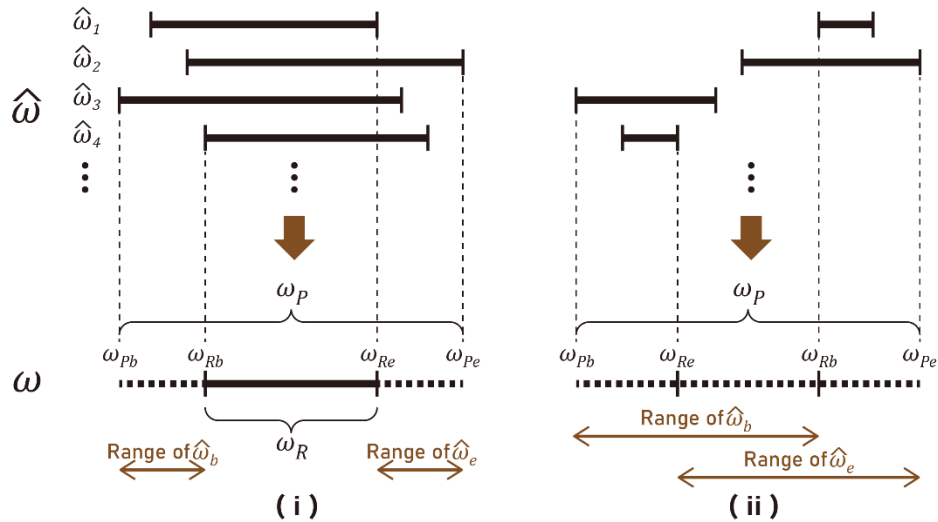

**Fig 1** Basic concept of the uncertain time interval. An uncertain time interval ($\omega$) is a set of determinate time intervals ($\hat{\omega}$). Ranges of the beginning and end points are separate in the left example, but they overlap in the right example (i). Consequently, possible time intervals ($\omega_P$) exist in both uncertain time intervals, whereas a reliable time interval ($\omega_R$) exists only in the uncertain time interval of the left example (ii).

When $\omega_{Rb}$ is less than or equal to $\omega_{Re}$, that is, if the ranges of the beginning and end points are separate, a reliable time interval $\omega_R$ exists, and its beginning and end points are $\omega_{Rb}$ and $\omega_{Re}$, respectively. Conversely, when $\omega_{Rb}$ is greater than $\omega_{Re}$, that is, if the ranges of the beginning and end points overlap, a reliable time interval $\omega_R$ does not exist (Figure 1). The reliable time interval indicates the temporal range that is certainly included in the uncertain time interval $\omega$.

When the possible and reliable time intervals in an uncertain time interval are equal ($\omega_P - \omega_R = \emptyset$), the beginning and end points are not temporal ranges but time instants; in this case, the uncertain time interval contains only one determinate time interval. This state indicates that the time interval is not uncertain.

## 3.2 Show novelty and usefulness definitely

### (1) Determinate time intervals

Before considering relations between uncertain time intervals, we consider the representation of relations between determinate time intervals. In this study, we classify relations between time intervals according to Allen's time interval relations (Figure 2).

A condition where a relation between a pair of determinate time intervals $\hat{a}$ and $\hat{b}$ is an Allen relation $\lambda$ is represented as Equation (5).

$$P_\lambda(\hat{a}, \hat{b}) \tag{5}$$

For example, let the beginning and end points of the determinate time interval $\hat{a}$ be $\hat{a}_b$ and $\hat{a}_e$, respectively; similarly, let the beginning and end points of $\hat{b}$ be $\hat{b}_b$ and $\hat{b}_e$, respectively. The condition for the relation between them to be Allen's **contains** relation (Figure 2) is represented as follows:

$$P_{\text{contains}}(\hat{a}, \hat{b}) = \hat{a}_b < \hat{b}_b \wedge \hat{a}_e > \hat{b}_e \ .$$

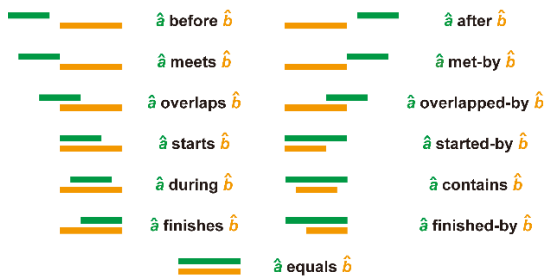

**Fig. 2** Allen's 13 relations between time intervals [13]. Inverting the time intervals of the left column relation results in the right column relation (e.g., $\hat{a}$ **before** $\hat{b}$ = $\hat{b}$ **after** $\hat{a}$).

### (2) Uncertain time intervals

Determining whether a relation between a pair of uncertain time intervals is an Allen relation $\lambda$ can be verified by applying a condition represented by Equation (5) to all elements in the Cartesian product of the uncertain time intervals ($a \times b$). For $a \times b$, which is a combination of the uncertain time intervals $a \ni \hat{a}$ and $b \ni \hat{b}$, there are three possible results: 1) all combinations of $(\hat{a}, \hat{b})$ satisfy the condition; 2) no combinations of $(\hat{a}, \hat{b})$ satisfy the condition; 3) some of the combinations of $(\hat{a}, \hat{b})$ satisfy the condition but others may not.

Here, we define the state where all combinations of $(\hat{a}, \hat{b})$ satisfy the condition for Allen relation $\lambda$ as a *reliable relation* (Equation 6).

$$\forall \hat{a} \forall \hat{b} P_\lambda(\hat{a}, \hat{b}) \tag{6}$$

A pair of time intervals in a reliable relation is always in Allen relation $\lambda$, even though both of the time intervals are uncertain. In an example of the reliable relation of Figure 3, the possible time interval $b_P$ of $b$ is located inside the reliable time interval $a_R$ of $a$. This state indicates that the temporal range where it is certain to be in $a$ contains the whole temporal range where it is possible to be in $b$, and, therefore, $(\hat{a}, \hat{b})$ always satisfies Allen's **contains** relation.

In contrast, we define a state where no combinations of $(\hat{a}, \hat{b})$ satisfy the condition for Allen relation $\lambda$ as an *impossible relation* (Equation 7).

$$\forall \hat{a} \forall \hat{b} \neg P_\lambda(\hat{a}, \hat{b}) \tag{7}$$

When a pair of time intervals is in the impossible relation, it is certain that they are never in Allen relation $\lambda$, even though both of the time intervals are uncertain. In an example of the impossible relation of Figure 3, a part of the reliable time interval $b_R$ is located outside the possible time interval $a_P$. This state indicates that a part of the temporal range where it is certain to be in $b$ is already located outside the temporal range where it is possible to be in $a$, and, therefore, $(\hat{a}, \hat{b})$ never satisfies Allen's **contains** relation.

In addition, as mentioned above, there is the state in which some combinations of $(\hat{a}, \hat{b})$ satisfy and others do not satisfy the condition to be in Allen relation $\lambda$. In this study, we define this state as a *possible relation*, represented by Equation (8).

$$\exists \hat{a} \exists \hat{b} P_\lambda(\hat{a}, \hat{b}) = \overline{\forall \hat{a} \forall \hat{b} \neg P_\lambda(\hat{a}, \hat{b})} \tag{8}$$

In an example of the possible relation of Figure 3, a part of the possible time interval $b_P$ is located outside the reliable time interval $a_R$. This state indicates that while there are some $(\hat{a}, \hat{b})$ that are in Allen's **contains** relation, some other $(\hat{a}, \hat{b})$ are in the other Allen relations (e.g., the **started-by** or **overlapped-by** relations).

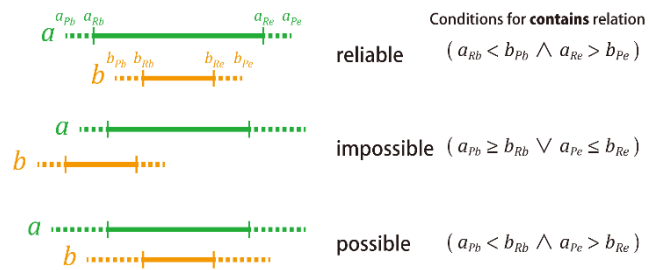

**Fig. 3** Three states (reliable, impossible, and possible) in relations between uncertain time intervals. These examples show states in Allen's **contains** relation and conditions to be in each state. Solid lines indicate reliable time intervals. Temporal ranges of the possible time interval outside the reliable time interval are indicated by dotted lines.

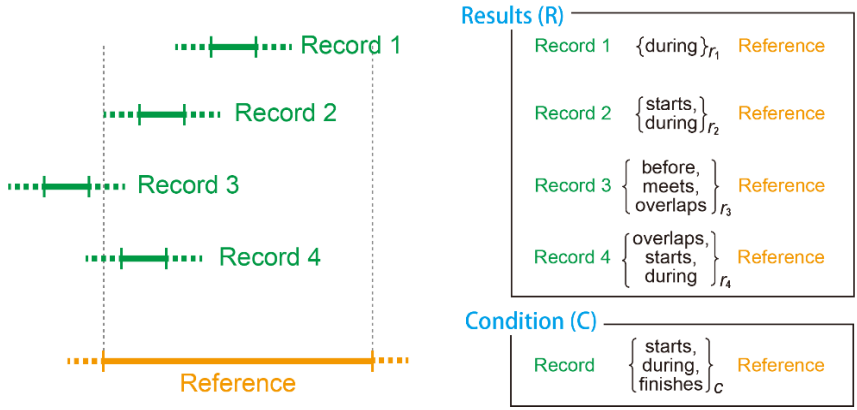

**Fig. 4** An example of the retrieval of uncertain time intervals according to a given condition. It is assumed that both the periods of records and the reference period are uncertain. The retrieval condition is given as a set of relations that should be *possible* between the periods of records and the reference period.

The state represented by Equation (8) includes the state of the reliable relation, in which all $(\hat{a}, \hat{b})$ satisfy the condition, in addition to the state in which some $(\hat{a}, \hat{b})$ satisfy it and others do not. This distinction—whereby the state of the possible relation includes the state of the reliable relation—is motivated by computational reasons. Equations (7) and (8) complement each other: if one is true, the other must be false. This means that we can easily know the result of testing for the impossible relation from the result of testing for the possible relation. Sekino (2019) [12] has derived conditions of these three states for all 13 Allen relations (Appendix Table1), but the conditions for the impossible relation are not shown, because we can easily derive them from the conditions for the possible relation.

### 3.3 Combination of Allen relations

In the previous section, we considered the logic to examine a specified Allen relation $\lambda$ individually; in this section, we consider a logic to examine the relations between uncertain time intervals for a set of Allen relations ($\Lambda \ni \lambda$).

#### (1) States of Allen Relations between Uncertain Time Intervals

Because Allen's 13 relations cover all relations between a pair of intervals, there is no state in which a pair of uncertain time intervals is *impossible* for all 13 Allen relations. Because a relation between a pair of time intervals always satisfies only one Allen relation, when a pair of uncertain time intervals is *reliable* for one Allen relation, it is always *impossible* for the other 12 Allen relations. Additionally, because the state of the *possible* relation includes the state of the *reliable* relation, when a pair of uncertain time intervals is *reliable* for an Allen relation, it is also *possible* for the relation. A pair of uncertain time intervals is *possible* for one or more Allen relations, and may even be *possible* for all 13 Allen relations. These features of the three states of relations mean that we can know the three states of every Allen relation by examining whether it is *possible* for each of the 13 Allen relations. Naturally, Allen relations that are not *possible* are *impossible*. When a pair of uncertain time intervals is *possible* only for one Allen relation, the other 12 Allen relations are *impossible*. It means that the pair is fixed at the one Allen relation, and is *reliable* for that Allen relation as a result.

#### (2) Retrievals using Uncertain Time Intervals

The retrieval of records with uncertain time intervals is a process that examines relations between the time intervals of the records and a reference time interval given as a retrieval condition. For example, for a retrieval of "paintings created during the Renaissance," we examine the relations between the creation period of paintings in a record and the period of the Renaissance, which is the reference period for the retrieval. If the relations satisfied include Allen's **starts**, **during**, or **finishes** relation, the record matches the retrieval condition. In the process of examining the relation according to a given retrieval condition, all 13 Allen relations between the time intervals of the record and of the reference period are examined, to determine whether they are *possible*; then the results are compared with the retrieval condition. For example, record 1 in Figure 4 is *possible* only for Allen's **during** relation. Because the **during** relation is contained in the retrieval condition (C), record 1 matches the condition. Record 2 in Figure 4 is *possible* for Allen's **starts** and **during** relations. Because both **starts** and **during** relations are contained in the retrieval condition (C), record 2 also matches the condition. Record 3 is *possible* for the **before**, **meets**, and **overlaps** relations, but none of them is contained in the retrieval condition (C). This indicates that the record 3 does not match the condition. How should we treat record 4, which is *possible* for the **overlaps**, **starts**, and **during** relations? Although the **starts** and the **during** relations are contained in the retrieval condition (C), the **overlaps** relation is not. This state means that there is a mixture of determinate time intervals that match the condition and determinate time intervals that do not match, in the uncertain time interval representing the record period. This is an example of the possible relation represented by Equation (8), and indicates that the record may match the condition, but it is not certain. Therefore, in the same manner as the relations between uncertain time intervals, we determine that records 1 and 2 are *reliable* matches, record 3 is an *impossible* match, and record 4 is a *possible* match for the retrieval condition.

This process to examine relations between uncertain time intervals according to a given retrieval condition can be generalized as follows. Let $c$ be a set of expected Allen relations given as a retrieval condition (Equation 9).

$$c = \{\lambda | \lambda \in \Lambda \wedge \lambda \text{ are expected relations}\} \qquad (9)$$

Let $a$ and $b$ be the uncertain time intervals representing the period of a record and the reference period for the retrieval,

respectively, and let $r$ be a set of *possible* Allen relations between $a$ and $b$ (Equation 10).

$$r = \{\lambda | \lambda \in \Lambda \wedge \exists \hat{a} \exists \hat{b} P_\lambda(\hat{a}, \hat{b})\} \tag{10}$$

Because the examination according to the condition $c$ is the same process as validation of whether $r$ is included in $c$, the examination $S(c, r)$ can be represented as Equation (11).

$$S(c, r) := \begin{cases} c \cap r = r & \Rightarrow & \text{Reliable} \\ c \cap r = \emptyset & \Rightarrow & \text{Impossible} \\ c \cap r \neq \emptyset & \Rightarrow & \text{Possible} \end{cases} \tag{11}$$

### 3.4 Time instant

A time instant, which is a value on a temporal axis, and a time interval, which is a set of time instants, are conceptually different entities. Nevertheless, in the actual processing, a time interval whose duration is 0 can be used instead of a time instant. For a determinate time interval, a time interval with a duration of 0 is one in which the beginning and end points ($\hat{\omega}_b$ and $\hat{\omega}_e$) are equal. For an uncertain time interval, a time interval with a duration of 0 is one in which the four values representing the ranges of the beginning and end points ($\omega_{Pb}$, $\omega_{Rb}$, $\omega_{Rb}$, and $\omega_{Pe}$) have the same value.

The concept of uncertain time intervals can be applied to represent uncertain time instants. Because the beginning and end points of a time interval with a duration of 0 are equal, an uncertain time instant is represented as an uncertain time interval whose ranges of beginning ($\omega_{Pb} \leq \omega_b \leq \omega_{Rb}$) and end ($\omega_{Re} \leq \omega_e \leq \omega_{Pe}$) points are the same (i.e., $\omega_{Pb} = \omega_{Re}, \omega_{Rb} = \omega_{Pe}$). Determinate time instants ($\{\hat{\omega} \in \omega | \hat{\omega}_b = \hat{\omega}_e\}$) contained in this uncertain time instant exist in the whole range between $\omega_{Pb} = \omega_{Re}$ and $\omega_{Rb} = \omega_{Pe}$. However, this uncertain time instant represented as an uncertain time interval also contains determinate time intervals whose duration is greater than 0 ($\{\hat{\omega} \in \omega | \hat{\omega}_b < \hat{\omega}_e\}$). This means that we cannot distinguish an uncertain time instant from an uncertain time interval. If a process needs only time instants or only time intervals with duration greater than 0, it is necessary to restrict them in the process based on the duration.

In this study, relations between two time instants or between a time instant and a time interval are defined according to the concept of OWL-Time [1]. It is defined that a pair of time instants can satisfy Allen's **before**, **equals**, or **after** relation. In the same manner, it can be defined that a time instant and a time interval can satisfy Allen's **before**, **starts**, **during**, **finishes**, or **after** relation. A time instant and a time interval can satisfy Allen's **meets**, **overlaps**, and **starts** relations simultaneously, when the time instant is located at the beginning point of the time interval. However, there must be only one Allen relation that is satisfied. To satisfy this restriction, this relation between a time instant and a time interval is defined to be Allen's **start** relation. In OWL-Time, an **in** relation was newly defined as a relation combining Allen's **starts**, **during**, and **finishes** relations. Therefore, when a time instant is located at the beginning point of a time interval, they satisfy the **in** relation as well as the **start** relation. In the same manner, it is defined that when a time instant is located at the end point of the time interval, they satisfy Allen's **finishes** relation, even though they could also satisfy Allen's **overlapped-by** and **met-by** relations. Inverse relations (between a time interval and a time instant) are Allen's **before**, **started-by**, **contains**, **finished-by**, and **after** relations. Conditions for the three states (*reliable*, *impossible*, and *possible* relations) between uncertain time intervals (Appendix Table 1) are also applicable to time instants.

## 4. Resources to construct data about uncertain time intervals

### 4.1 Data description

To describe uncertain time intervals as data, the four time instants ($\omega_{Pb}$, $\omega_{Rb}$, $\omega_{Re}$, and $\omega_{Pe}$), representing the ranges of the beginning and end points, are specified. Character strings according to ISO 8601, values of datetime type of a database system, or other values to represent a position on a temporal axis are available to specify the time instants. Additionally, the ranges of the beginning and end points can be specified by other time intervals. For example, when a range of the beginning point is specified by a calendar date, the beginning point of the described uncertain time interval is represented as the range from 00:00 to 24:00 on the given date. Other time intervals are also available to specify the range of the beginning and end points instead of calendar dates. For determinate time intervals in which both boundaries are fixed, the values of the beginning and end points are used to specify the range. For uncertain time intervals, the values of the beginning and end points of their possible time interval is used to specify the range.

Whether time instants or time intervals are used to describe an uncertain time interval, procedures are required to obtain the four values of time instant that represent the ranges of the boundaries, and to compare them, to examine the relation between the uncertain time intervals. In this respect, it is effective to use Julian dates as values of time instants. The Julian date is the total day count from noon on January 1st, BCE 4713 [14], and is widely used in various scientific fields such as astronomy. Because the positions on the temporal axis are specified as real numbers, it is easy to compare time instants represented by the Julian date. There are many services on the web to convert dates between various types of calendars, including conversion from the Gregorian calendar to Julian dates (for example, the HuTime Calendar Conversion Service [15]).

It is effective to use the HuTime Calendar linked open data (LOD) [16] when the data are described using resource description framework (RDF), applying semantic web technology. The HuTime Calendar LOD provides information about calendar periods, such as dates, months, years, and eras as RDF resources, which are available to specify the ranges of boundaries in uncertain time intervals. For example, the calendar month of April 2019 can be specified as a resource using its URI (http://datetime.hutime.org/calendar/101.1/month/2019-04). All resources of calendar periods provided by the HuTime Calendar

LOD are linked to Julian date values representing the beginning and end of the calendar period. Therefore, values of time instants to examine relations between uncertain time intervals can easily be obtained when the HuTime Calendar LOD is used to specify the ranges of boundaries in uncertain time intervals (Figure 5).

```
@prefix time: <http://www.w3.org/2006/time#>
@prefix hutime: <http://resource.hutime.org/ontology/>
@prefix hcal: <http://datetime.hutime.org/calendar/>

ex:Term1   a
           hutime:hasPossibleBeginning  [
                                 a
                                 time:inXSDDate
                                 "2018-01-01"^^xsd:date ; ] ;
           hutime:hasReliableBeginning  [
                                 a
                                 time:inXSDDate
                                 "2018-03-01"^^xsd:date ; ] ;
           hutime:hasReliableJdEnd   2458300.5^^xsd:double ;
           hutime:hasPossibleJdEnd   2458300.5^^xsd:double ;

ex:Term2   a                          hutime:UncertainTimeInterval ;
           hutime:hasRangeOfBeginning hcal:101.1/year/2019 ;
           hutime:hasRangeOfEnd       hcal:101.1/year/2019 ;

ex:Term3   a                      time:ProperInterval ;
           time:hasBeginning      [
                                 a
                                 time:inXSDDate
                                 "2020-07-01"^^xsd:date ; ] ;
           time:hasEnd            [
                                 a
                                 time:inXSDDate
                                 "2021-01-01"^^xsd:date ; ] ;

ex:Term4   a                          hutime:UncertainTimeInterval ;
           hutime:hasRangeOfBeginning ex:Term1 ;
           hutime:hasRangeOfEnd       ex:Term3 .
```

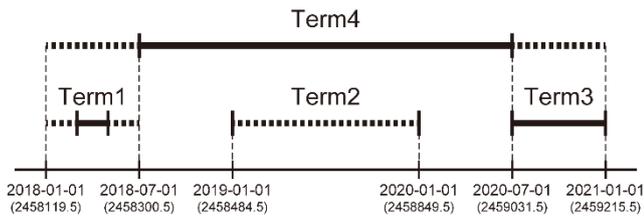

**Fig. 5** Example data describing uncertain time intervals according to the HuTime Ontology and schematic image of the data structure. Beginning range of Term 4 is specified by Term 1 using a property of hutime:hasRangeOfBeginning. The beginning and end points of the range are the beginning and end points of the possible time interval (shown by dotted line) of Term1, respectively. End rang of Term 4 is also specified by Term3 in the same way as the beginning range. This process of creating Term 4 shows the fact that uncertain time intervals can be defined, recursively. This means that Term 4 is also available to specify the beginning or end ranges of the other uncertain time interval.

## 4.2 Ontology and vocabularies

In this study, an ontology (HuTime Ontology) [17], including vocabularies to describe uncertain time intervals as data, was created (http://resource.hutime.org/ontology/) (Figure 6). The HuTime ontology is designed to be compatible with OWL-Time, and is also an extension of OWL-Time.

In OWL-Time, the beginning and end points of time intervals are defined as fixed time instants; therefore, uncertain time intervals, whose beginning and end points are temporal ranges, cannot be processed in the OWL-Time framework. To solve this problem, a new class to represent uncertain time intervals was defined in this study, associated with time intervals defined in OWL-Time. Actual time intervals are defined as

time:ProperInterval in OWL-Time (prefix "time:" indicates http://www.w3c.org/2006/time#, which is a part of the URI). A time interval represented by time:ProperInterval is equivalent to a special form of uncertain time interval, of which the range durations of both boundaries are 0. Therefore, a new class hutime:UncertainTimeInterval, representing uncertain time intervals, was defined as a superclass of time:ProperInterval (prefix "hutime:" indicates http://resource.hutime.org/ontology/, which is a part of the URI). Additionally, vocabularies to describe uncertain time intervals were defined as properties of hutime:UncertainTimeInterval. Four properties (for example, hutime:hasPossibleBeginning) were defined to specify the four time instants that determine an uncertain time interval. The domain and range of those properties are hutime:UncertainTimeInterval and time:TemporalEntity, respectively. Because time:TemporalEntity is the top-level entity that includes all types of time entities in OWL-Time, all types of resources according to OWL-Time are available to describe an uncertain time interval (the beginning range of Term 1 in Figure 5). Furthermore, properties to represent the four time instants of an uncertain time interval using Julian dates were also defined (for example, hutime: hasPossibleJdBeginning). Ranges of these properties are numbers of the double type. Values to examine relations between uncertain time intervals can easily be obtained through these properties (the end range of Term 1 in Figure 5). Properties hutime:hasRangeOfBeginning and hutime:hasRangeOfEnd are defined to specify the ranges of the beginning and end points in an uncertain time interval using other time intervals. The domain and range of these two properties are the same—hutime:UncertainTimeInterval—and reflect the feature of uncertain time intervals that they can be created recursively (the beginning point of Term 4 in Figure 5).

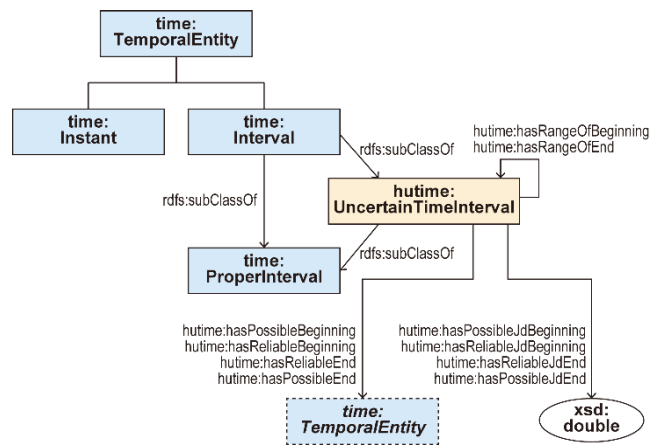

**Fig. 6** Classes and properties of HuTime Ontology [17] and its relation to OWL-Time. The four values ($\omega_{Pb}$, $\omega_{Rb}$, $\omega_{Re}$, and $\omega_{Pe}$) to describe an uncertain time interval can be specified by resources of time:TemporalEntity or Julian dates. Ranges of the beginning and the end points of the uncertain time interval also can be specified by other resources of hutime:UncertainTimeIntervals.

Because the HuTime Ontology was designed according to OWL-Time, there are compatibilities between them. For example, because time:ProperInterval is a subclass of

hutime:UncertainTimeInterval, time intervals of time:ProperInterval are available in analyses and retrievals using uncertain time intervals. For the same reason, time intervals of time:ProperInterval are available to specify the ranges of the beginning and end points of uncertain time intervals (the end range of Term 4 in Figure 5). However, to process an uncertain time interval (hutime:UncertainTimeInterval) as a time interval (time:ProperInterval), it is necessary to convert between them. Usually, the possible time intervals of uncertain time intervals are appropriate for use as the time intervals of time:ProperInterval. The HuTime Ontology and OWL-Time are just conceptual models describing relationships between classes and properties; therefore, the actual functions for data conversion depend on the implementation of each system.

The HuTime Calendar LOD [16] provides resources according to the HuTime Ontology. Calendar periods such as dates, months, and years are described as instances of hutime:UncertainTimeInterval, and the beginning and end points are given as Julian dates, through properties such as hutime:hasPossibleJdBeginning. Therefore, when these calendar period resources are used to specify ranges of boundaries in uncertain time intervals (Term 2 in Figure 5), values required for retrievals and analyses of them can be easily obtained. We expect SPARQL functions [17] and Web API, corresponding to HuTime Ontology, to be implemented in future.

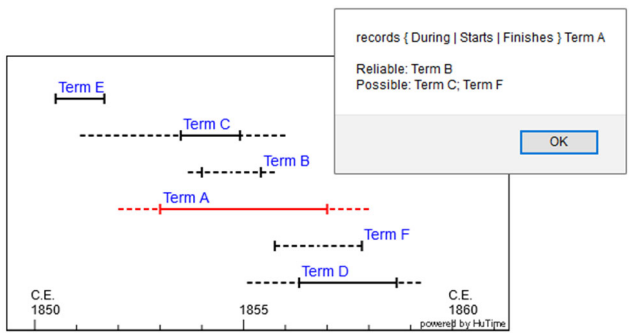

**Fig. 7** Uncertain time intervals displayed on a web application HuTime. A retrieval result is shown in a popup window. These functions are an experimental implementation of the logics constructed in this study.

## 5. Experimental implementation

Based on the logics in Section 3, functions of visualization and retrieval for uncertain time intervals were implemented in a web application HuTime (Figure 7). HuTime is a time information system developed according to a design concept similar to a geographic information system (GIS): it displays and analyzes temporal data on timelines and time-series charts [19]. Uncertain time intervals are displayed on HuTime as solid and dotted lines, indicating the reliable and possible time intervals, in the same way as Figure 3. HuTime accepts temporal data of various types of calendars, including data according to ISO 8601, and converts them to Julian dates when the data are loaded. Therefore, HuTime can efficiently retrieve and analyze uncertain time intervals. This experimental implementation indicates that the logics constructed in this study are implementable.

## 6. Application

Fig. 8 shows historical periods and person's life spans represented as uncertain time intervals. A boundary between the early and late modern period shown in Fig. 8 is vague, because the early modern period did not suddenly transition to the late modern period on a certain day. Therefore, these historical periods are represented as uncertain time intervals in Fig. 8. Since the boundary of these historical periods is often indicated by the industrial revolution, in European history, the end range of the early modern period and the beginning range of the late modern period are indicated by the period of the industrial revolution in Fig. 8. As a result, a vague boundary between the early and late modern periods is appropriately shown using the logic of this study.

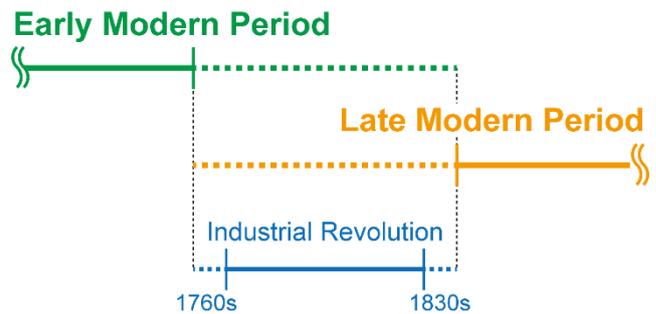

```
@prefix hutime: <http://resource.hutime.org/ontology/>
@prefix hcal: <http://datetime.hutime.org/calendar/>

ex:EarlyModernPeriod        a                        hutime:UncertainTimeInterval ;
                            hutime:hasRangeOfBeginning ex:Renaissance ;
                            hutime:hasRangeOfEnd       ex:IndustrialRevolution ;

ex:LateModernPeriod         a                        hutime:UncertainTimeInterval ;
                            hutime:hasRangeOfBeginning ex:IndustrialRevolution ;
                            hutime:hasRangeOfEnd       ex:WorldWarII ;

ex:IndustrialRevolution     a                        hutime:UncertainTimeInterval ;
                            hutime:hasRangeOfBeginning hcal:101.2/10y/1760s ;
                            hutime:hasRangeOfEnd       hcal:101.2/10y/1830s ;

ex:JamesWatt                a                        hutime:UncertainTimeInterval ;
                            hutime:hasRangeOfBeginning hcal:101.2/date/1736-01-19 ;
                            hutime:hasRangeOfEnd       hcal:101.2/date/1819-08-25 ;

ex:GeorgeStephenson         a                        hutime:UncertainTimeInterval ;
                            hutime:hasRangeOfBeginning hcal:101.2/date/1781-06-09 ;
                            hutime:hasRangeOfEnd       hcal:101.2/date/1848-08-12 ;
```

**Fig. 8** An application example of the logics constructed in this study. Historical periods and person's life span are represented as uncertain time intervals.

The logic of this study also can be applied to extract a person who was alive during these periods. If a person was alive during a certain period, the relationship between the life span and the period can be shown as Allen **overlaps**, **starts**, **during**, **finishes**, **overlapped-by**, **started-by**, **contains**, **finished-by** and **equals** relations (When the life span is enough shorter than the historical period, Allen **started-by**, **contains**, **finished-by**, and **equals** relations can be ignored). These relations are corresponding to $c$ in Equation (11).

Here, we consider a query of "Was James Watt (1736-1819) alive during the early modern period?". Possible Allen relations between his life span and the reference period are **during**, **finishes**, and **overlapped-by**, and are corresponding to $r$ in Equation (11). As a result, the answer of the query is *reliable* according to Equation (11). On the other hand, the answer of a query "Was James Watt alive during the late modern period?" is *possible*, because Allen relations between his life span and the period (i.e., $r$ in Equation (11)) are **before**, **meets**, and **overlaps**. In the same manner, we can obtain query results that George Stephenson (1781-1848) was *reliably* alive during the late modern period and was *possibly* alive during the early modern period.

These results are different from results referencing a period represented by a conventional way which strictly separates periods at a time instant. When the early and the late modern periods are separated at the end of the industrial revolution (the end of 1839) according to the conventional way, the result indicates that James Watt was alive only during the early modern period. While, when these periods are separated at the beginning of the industrial revolution (the beginning of 1760), the result indicates that he was alive during both the early and the late modern period. However, the result does not have information about certainty indicated by *reliable* and *possible*. Therefore, these two periods in the result will be processed in the same manner in spite of a possibility that they may differ in certainty. In contrast, the logic of this study can consider vague of the boundary between the periods, and allows to obtain appropriate results including the certainty.

## 7. Discussion

The logics and resources in this study can also be applied for various kinds of analysis, even though they have been constructed mainly to realize retrievals. The relations between the time periods when two events happened can be used as evidence of causality between them. The overlap between the periods when two objects existed is a clue to the interaction between them (for example, a relation between persons and documents, to identify the authors). These temporal information analyses can be realized using the logics constructed and implemented in this study.

The HuTime Ontology constructed in this study has the feature that the uncertain time intervals can be defined recursively. This means that the ranges of the beginning and end points of an uncertain time interval can be specified using other uncertain time intervals, according to the HuTime Ontology. For example, when "Industrial Revolution" and "Great Depression" are defined as periods according to the HuTime ontology, it is possible to define a new period that indicates the "period between the Industrial Revolution and the Great Depression," and the new period can be used not only for retrievals but also for data description. If a thesaurus of information about periods is constructed according to the HuTime Ontology, it would be an effective research resource for data description and the analysis of temporal information.

## 8. Conclusion

In this study, new logics to process uncertain time intervals were constructed. Uncertain time intervals are described, specifying ranges of the beginning and end points, and their nature is explained by the possible and reliable time intervals. States of relations between a pair of uncertain time intervals can be classified into *reliable*, *impossible*, and *possible*, for each Allen relation. Applying these logics, a logic to retrieve uncertain time intervals according to a given condition was constructed. These logics are simpler and clearer than conventional logics, and therefore are appropriate for practical implementation.

Based on these logics, the HuTime Ontology to describe data of uncertain time intervals was constructed. Because the HuTime Ontology was designed according to OWL-Time, which is widely used, it has a high affinity with existing data.

Finally, functions to realize the logics constructed in this study were experimentally implemented in an existing web application HuTime, to prove that the logics were implementable. These implemented functions have already been released as a library, and are the first implementations that have functions to retrieve and analyze uncertain time intervals, as well as to describe them.

The results of this study will make it possible to use temporal information that has not previously been used because of uncertain time periods, and will promote the unification of procedures and criteria to process uncertain time periods that differ between databases. It is expected that more temporal data will be constructed and released utilizing the results of this study.

**Acknowledgments**  This study is mainly supported by JSPS KAKENHI "Semantic chronology – Construction of infrastructure for visualization and utilization of knowledge along temporal axis" (15H01723) and partially supported by JSPS KAKENHI JSPS KAKENHI 16H01897, 16H01830, 16H02918, 17H00773, and 18H03588.

# Appendix

Table 1 shows whether each of the 13 Allen relations is *reliable* or *possible* [12] between two uncertain time intervals $a$ and $b$, where $a$ is characterized by $a_{Pb}, a_{Pe}, a_{Rb}, a_{Re}$, and likewise for $b$.

**Table 1**. Conditions for each Allen relation to be a *reliable* or *possible* relation [12].

| Relation | Reliable | Possible |
|---|---|---|
| before | $a_{Pe} < b_{Pb}$ | $a_{Re} < b_{Rb}$ |
| after | $a_{Pb} > b_{Pe}$ | $a_{Rb} > b_{Re}$ |
| during | $a_{Pb} > b_{Rb} \wedge a_{Pe} < b_{Re}$ | $a_{Rb} > b_{Pb} \wedge a_{Re} < b_{Pe} \wedge b_{Pb} < b_{Pe}$ |
| contains | $a_{Rb} < b_{Pb} \wedge a_{Re} > b_{Pe}$ | $a_{Pb} < b_{Rb} \wedge a_{Pe} > b_{Re} \wedge a_{Pb} < a_{Pe}$ |
| overlaps | $a_{Rb} < b_{Pb} \wedge a_{Re} > b_{Rb} \wedge a_{Pe} < b_{Re}$ | $a_{Pb} < b_{Rb} \wedge a_{Pe} > b_{Pb} \wedge a_{Re} < b_{Pe}$ $\wedge a_{Pb} < a_{Pe} \wedge b_{Pb} < b_{Pe}$ |
| overlapped-by | $a_{Pb} > b_{Rb} \wedge a_{Rb} < b_{Re} \wedge a_{Re} > b_{Pe}$ | $a_{Rb} > b_{Pb} \wedge a_{Pb} < b_{Pe} \wedge a_{Pe} > b_{Re}$ $\wedge a_{Pb} < a_{Pe} \wedge b_{Pb} < b_{Pe}$ |
| meets | $a_{Re} = a_{Pe} = b_{Pb} = b_{Rb}$ $\wedge a_{Rb} < b_{Pb} \wedge a_{Pe} < b_{Re}$ | $a_{Re} \le b_{Rb} \wedge a_{Pe} \ge b_{Pb} \wedge a_{Pb} < b_{Rb}$ $\wedge a_{Re} < b_{Pe} \wedge a_{Pb} < a_{Pe} \wedge b_{Pb} < b_{Pe}$ |
| met-by | $a_{Pb} = a_{Rb} = b_{Re} = b_{Pe}$ $\wedge a_{Pb} > b_{Rb} \wedge a_{Re} > b_{Pe}$ | $a_{Pb} \le b_{Pe} \wedge a_{Rb} \ge b_{Re} \wedge a_{Rb} > b_{Pb}$ $\wedge a_{Pb} > b_{Re} \wedge a_{Pb} < a_{Pe} \wedge b_{Pb} < b_{Pe}$ |
| starts | $a_{Pb} = a_{Rb} = b_{Pb} = b_{Rb} \wedge a_{Pe} < b_{Re}$ | $a_{Pb} \le b_{Rb} \wedge a_{Rb} \ge b_{Pb} \wedge a_{Re} < b_{Pe} \wedge b_{Pb} < b_{Pe}$ |
| started-by | $a_{Pb} = a_{Rb} = b_{Pb} = b_{Rb} \wedge a_{Re} > b_{Pe}$ | $a_{Pb} \le b_{Rb} \wedge a_{Rb} \ge b_{Pb} \wedge a_{Pe} > b_{Re} \wedge a_{Pb} < a_{Pe}$ |
| finishes | $a_{Pb} > b_{Rb} \wedge a_{Re} = a_{Pe} = b_{Re} = b_{Pe}$ | $a_{Rb} > b_{Pb} \wedge a_{Re} \le b_{Pe} \wedge a_{Pe} \ge b_{Re} \wedge b_{Pb} < b_{Pe}$ |
| finished-by | $a_{Rb} < b_{Pb} \wedge a_{Re} = a_{Pe} = b_{Re} = b_{Pe}$ | $a_{Pb} < b_{Rb} \wedge a_{Re} \le b_{Pe} \wedge a_{Pe} \ge b_{Re} \wedge a_{Pb} < a_{Pe}$ |
| equals | $a_{Pb} = a_{Rb} = b_{Pb} = b_{Rb} \wedge a_{Re} = a_{Pe} = b_{Re} = b_{Pe}$ | $a_{Pb} \le b_{Rb} \wedge a_{Rb} \ge b_{Pb} \wedge a_{Re} \le b_{Pe} \wedge a_{Pe} \ge b_{Re}$ |